\documentclass[prl,aps,twocolumn,amssymb,floats,floatfix]{revtex4}

\usepackage{epsfig}

\newcommand{\fpi}{f_\pi}
\newcommand{\mpi}{m_\pi}
\newcommand{\non}{\nonumber}
\newcommand{\figsize}{6cm}

\begin{document}

\preprint{
\vbox{
\hbox{ADP-01-49/T481, LTH-528}
}}

\title{Chiral Properties of Quenched and Full QCD}

\author{R. D. Young$^1$}
\author{D. B. Leinweber$^1$}
\author{A. W. Thomas$^1$}
\author{S. V. Wright$^{1,2}$}

\affiliation{$^1$Special Research Centre for the
                 Subatomic Structure of Matter,
                 and Department of Physics and Mathematical Physics,
                 University of Adelaide, Adelaide SA 5005,
                 Australia \\
             $^2$Division of Theoretical Physics,
                 Department of Mathematical Sciences, 
                 University of Liverpool,
                 Liverpool L69 3BX, U.K.}

\date{\today}
\begin{abstract}
We extend a technique for the chiral extrapolation of hadron masses 
calculated with dynamical fermions to those generated by quenched 
simulations. The method ensures the
correct leading and next-to-leading non-analytic  
behaviour for either QCD or quenched QCD in the chiral limit, as well 
as the correct large quark mass behaviour. 
We find that the primary difference between quenched 
and dynamical baryon masses can be described by the meson loops which 
give rise to the different leading and next-to-leading non-analytic 
behaviour.
\end{abstract}

\maketitle
Modern computing facilities, combined with innovations in improved
actions for lattice QCD, mean that it is now possible to perform accurate
quenched QCD (QQCD) simulations at quite low quark masses
\cite{Bowler:1999ae,Kanaya:1998sd,Bernard:2001av,Zanotti:2001yb}. 
For simulations with dynamical 
fermions (full QCD) the situation is much more difficult, but there are
initial results at quark masses as low as 30 MeV 
\cite{Bernard:2001av,Aoki:1999ff}. The latter development has inspired 
studies of chiral extrapolation aimed at
using the full QCD data over a range of masses to reliably extract the
physical hadron mass. 

In general, effective field theories, such as
chiral perturbation theory, lead to divergent or asymptotic
expansions \cite{Dyson:1952tj,LeGuillou:1990nq}. While this raises doubts 
about the direct application of chiral perturbation theory to lattice 
data, studies of the mass dependence of hadron properties in 
QCD-inspired models 
\cite{magmom,Hackett-Jones:2000js,Detmold:2001jb},
as well as the exactly soluble Euler-Heisenberg problem \cite{Dunne:2001ip},
suggest that one can develop surprisingly accurate
extrapolation formulas, provided one builds in the correct behaviour in
{\it both}\ the small and large mass limits. For the nucleon ($N$) and
delta ($\Delta$) masses (and by extension all other baryons), 
Leinweber et al. \cite{Leinweber:1999ig}
have suggested an extrapolation method which ensures both the exact low
mass limit of chiral perturbation theory (technically its leading (LNA)
and next-to-leading non-analytic (NLNA) behaviour) and the heavy 
quark limit of heavy quark effective theory (HQET). The transition between the
chiral and heavy quark regimes is characterised by a mass scale
$\Lambda$, related to the inverse of the size of the pion cloud source. 
The rapid, non-analytic variation of
hadron properties, characteristic of chiral perturbation theory, is
rapidly suppressed once the pion Compton wavelength is smaller than this
size (i.e. $m_\pi > \Lambda$).

It is straightforward to extend the method of 
Ref. \cite{Leinweber:1999ig} to QQCD. One simply includes all the
Goldstone loops (including both $\pi$ and $\eta'$) which give rise to
the LNA and NLNA behaviour of quenched chiral perturbation theory (Q$\chi$PT) 
\cite{Bernard:1992mk,Labrenz:1996jy}. 
In principle, the parameters of the chiral Lagrangian are dependent on 
the number of dynamical fermion flavours, $N_f$, but are independent 
of the masses of the quarks. This is a celebrated feature of 
partially quenched chiral perturbation theory 
\cite{Sharpe:2000bc,Sharpe:2001fh}. The extent of the $N_f$ 
dependence is not well known, and a precise determination awaits 
dynamical fermion simulations with light dynamical quarks of 
varying number. 

Phenomenological investigations
\cite{Hackett-Jones:2000js,Leinweber:1993hj} of the role of the
pion cloud in hadronic charge radii indicate that results consistent
with experiment can be obtained by adding full-QCD chiral corrections
to the results of quenched simulations\cite{Leinweber:1991dv} at
moderate to heavy quark masses.  This suggests that the size of the
pion source is not changed dramatically in going from the quenched
approximation to full QCD, motivating the use of a common 
scale, $\Lambda$, in QQCD and QCD. We proceed to fit quenched 
and dynamical lattice data assuming negligible  $N_f$ dependence in 
the chiral parameters and $\Lambda$. The extent to which these 
assumptions hold can only be determined via further dynamical 
fermion calculations in the light quark regime.

By incorporating the chiral loops which give rise to the
LNA and NLNA behaviour in QCD and QQCD we find a remarkable agreement 
between the fit parameters of each simulation. This supports our 
hypothesis that 
the behaviour of the source of the pion cloud within baryonic systems 
behaves much the same in both QQCD and QCD. 
The differences between quenched and full QCD 
is primarily described by the differences in the associated 
chiral loops. 
Since the chiral corrections are expected to be
larger for $N$ and $\Delta$ than for others, this suggests that a
similar technique may be applicable to {\it all}\ baryons. 
This would enable quenched simulations to play a more 
valuable role, together with new experimental information from 
JLab and elsewhere, in deepening our understanding of baryon 
spectroscopy.

With regard to the properties of the $N$ and $\Delta$ we find 
a spectacular difference in QQCD. Whereas the extrapolation of 
the $N$ mass is essentially linear in the quark mass, the 
$\Delta$ exhibits some upward curvature in the quenched chiral 
limit. As a result, the $\Delta$ mass in QQCD is expected to be 
of the order 300--400 MeV above its mass in full QCD. The 
success of the extrapolation scheme also lends confidence to 
the interpretation of the $\Delta-N$ mass splitting as 
receiving a contribution of order 50 MeV from pion loops in 
full QCD and up to 250 MeV in QQCD \cite{tocome}. The residual 
splitting in full QCD would then be naturally ascribed to some 
shorter range mechanism, such as the traditional 
one-gluon-exchange \cite{Isgur:1999jv}.

The method for extrapolating baryon masses proposed 
by Leinweber~et~al.~\cite{Leinweber:1999ig} is to fit the
lattice data with the form:
\begin{equation}
M_B = \alpha_B + \beta_B \mpi^2 + \Sigma_B (\mpi, \Lambda) ,
\label{Eq:fullfit}
\end{equation}
where $\Sigma_B$ is the sum of those pion loop induced self-energies
which give rise to the LNA and NLNA behaviour of the mass, $M_B$.
In the case of the $N$ this is the sum of the processes 
$N\to N\pi\to N$ and $N\to \Delta\pi\to N$ 
, while for the $\Delta$ it involves 
$\Delta \to \Delta \pi\to \Delta$ 
and $\Delta \to N \pi \to \Delta$. 
In the heavy baryon limit, these four contributions 
($B\to B'\pi\to B$) can be summarised as: 
\begin{equation}
\sigma_{BB'}^\pi =  -\frac{3}{16\pi^2 \fpi^2} G_{BB'} 
\int_0^\infty dk \frac{k^4 u_{BB'}^2(k)}
{\omega(k) ( \omega_{BB'} + \omega(k) )},
\label{eq:fullSEpi}
\end{equation}
where $\omega(k)=\sqrt{k^2+\mpi^2}$ and $\omega_{BB'} = (M_{B'} - M_B)$, and
the constants $G_{BB'}$ are standard SU(6) couplings
\cite{Leinweber:1999ig}.

The factor $u(k)$, which acts as an ultraviolet regulator, may 
be interpreted physically as the Fourier transform of the source of the
pion field. Whatever choice is made, the form of these meson loop
contributions guarantees the exact LNA and NLNA structure of
chiral perturbation theory ($\chi$PT). Furthermore, 
such a form factor causes the self-energies to decrease
as $1/m_\pi^2$ for $m_\pi >> \Lambda$.
One commonly uses a dipole, 
$u(k)=(\Lambda^2 - \mu^2)^2/(\Lambda^2 + k^2)^2$, 
with $\mu$ the physical pion mass.

Quenched $\chi$PT is a low energy effective 
theory for quenched QCD \cite{Bernard:1992mk,Labrenz:1996jy}, analogous
to $\chi$PT for full QCD \cite{Bernard:1995dp}. 
Sea quark loops are formally removed from QCD by including a set of 
degenerate, bosonic quarks. 
These bosonic fields have the effect of cancelling the fermion 
determinant in the functional integration over the quark fields.
This gives a Lagrangian field theory which is 
equivalent to the quenched approximation simulated on the lattice. 
The low energy effective theory is then constructed using 
the symmetry groups of this Lagrangian.

A study of the chiral structure of baryon masses within the quenched 
approximation has been carried out by Labrenz and Sharpe 
\cite{Labrenz:1996jy}. The essential differences from full QCD are:
a) in the quenched theory the chiral coefficients differ 
from their standard values and b) new non-analytic structure is also 
introduced.
The leading order form of the baryon mass expansion about $\mpi=0$ is
\begin{eqnarray}
M_B & = & M_B^{(0)} + c_1^B \mpi + c_2^B \mpi^2  + c_3^B \mpi^3  \non \\
    &   & + c_4^B \mpi^4 + c_{4L}^B \mpi^4 \log \mpi + \ldots
\label{eq:chiexp}
\end{eqnarray}
where the coefficients of terms non-analytic in the quark mass are 
model-independent. Throughout we use couplings as given in 
Ref.~\cite{Labrenz:1996jy}. In addition, 
we have included an octet--decuplet mass 
splitting to explicitly give a value for $c_{4L}^B$ \cite{tocome}. 
We also stress that the term in $\mpi$ is absent in full QCD --- such a 
term being unique to the quenched case.
%

In fitting quenched data we wish to replicate the analysis for full
QCD while incorporating the known chiral structure of the quenched
theory.  The meson-loop, self-energy corrections to baryon masses can
be described in the same form as for full QCD. The effect of quenching
appears as a redefinition of the couplings in the loop diagrams in
order that they yield exactly the same LNA and NLNA structure as given
by Q$\chi$PT. For example, the analytic expressions for the pion-cloud
corrections to the masses of the $N$ and $\Delta$ have the same form
as the full QCD integrals (Eq.~\ref{eq:fullSEpi}) with redefined
quenched couplings. We refer to Ref.~\cite{tocome} for details.
Assuming a weak $N_f$ dependence of the chiral parameters, we describe
the quenched self-energies using the same tree level values of
$D=0.76$ and $F=0.50$ as in full QCD.

In addition to the usual pion loop contributions, QQCD
contains loop diagrams involving the flavour singlet $\eta'$ which 
also give rise to important non-analytic structure. Within full QCD 
such loops do not play a role in the chiral expansion because the 
$\eta'$ remains massive in the chiral limit.
On the other hand, in the quenched approximation the 
$\eta'$ is also a Goldstone boson \cite{Bernard:1992mk,Sharpe:1992ft}  
and the $\eta'$ propagator is exactly the same 
as that of the pion.

\begin{figure}
\begin{center}
{\epsfig{file=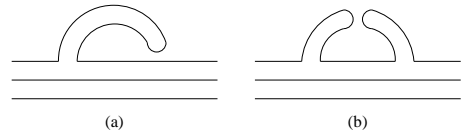, width=\figsize, angle=0}}
\caption{Quark flow diagrams of chiral $\eta'$ loop contributions
appearing in QQCD:  
(a) single hairpin, (b) double hairpin.
\label{fig:etaQF}}
\end{center}
\end{figure}

As a consequence there are two new chiral loop contributions 
unique to the quenched theory. The first of these, 
$\tilde \sigma_B^{\eta'(1)}$, corresponds to a single hairpin 
diagram such as that indicated in Fig.~\ref{fig:etaQF}(a).
This diagram is the source of the term proportional to $\mpi^3$ 
(involving the couplings $\gamma$ and $\gamma'$ \cite{Labrenz:1996jy}) 
in the chiral expansion Eq.~(\ref{eq:chiexp}). 
The structure of this diagram is exactly the same as the pion loop 
contribution where the internal baryon is degenerate with the 
external state. 
The second of these new $\eta'$ loop diagrams, 
$\tilde \sigma_{B}^{\eta'(2)}$, arises from the double 
hairpin vertex as pictured in Fig.~\ref{fig:etaQF}(b). This contribution 
is particularly interesting because it involves two Goldstone boson 
propagators and is therefore the source of the non-analytic term 
linear in $\mpi$.

\begin{figure}[tbh]
\begin{center}
{\epsfig{file=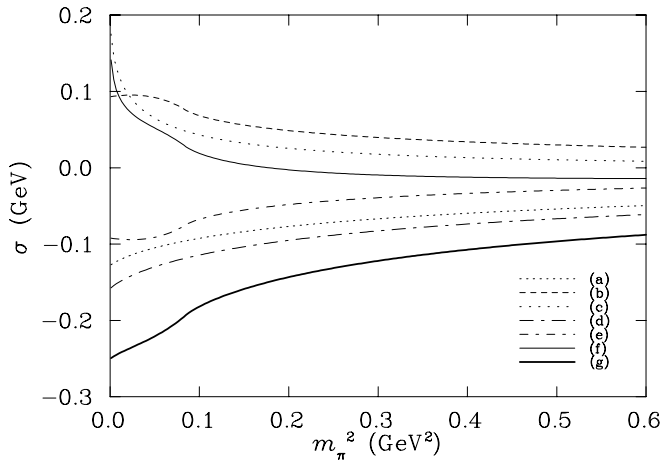, width=\figsize, angle=90}}
\caption{Various self-energy contributions to $M_\Delta$ for 
dipole mass, $\Lambda=0.8\, {\rm GeV}$. Shown are those of quenched QCD: 
(a)~$\tilde{\sigma}^\pi_{\Delta\Delta}$, 
(b)~$\tilde{\sigma}^\pi_{\Delta N}$, 
(c)~$\tilde{\sigma}_\Delta^{\eta'(2)}$;
and full QCD:
(d)~$\sigma^\pi_{\Delta\Delta}$,
(e)~$\sigma^\pi_{\Delta N}$;
and net totals:
(f)~$\tilde\Sigma_\Delta$,
(g)~$\Sigma_\Delta$.}
\label{fig:seD}
\end{center}
\end{figure}
The total meson loop contribution to the baryon self-energy within the 
quenched approximation is given by the sum of these four diagrams:
\begin{equation}
\tilde{\Sigma}_B = \tilde{\sigma}^\pi_{BB} + \tilde{\sigma}^\pi_{BB'} +
               \tilde{\sigma}_B^{\eta'(1)} + \tilde{\sigma}_B^{\eta'(2)}.
\end{equation}
As the resultant pion couplings in QQCD are quite a bit smaller than the
corresponding full QCD couplings, $\tilde \Sigma_B$ is smaller in
magnitude than $\Sigma_B$. We display the individual contributions 
to the $\Delta$ mass for both quenched and full QCD in Fig.~\ref{fig:seD}.
It is notable that $\tilde \sigma^\pi_{\Delta N}$ and 
$\tilde \sigma_{\Delta}^{\eta'(2)}$ are repulsive, with 
$\tilde \sigma_{\Delta}^{\eta'(1)}$ vanishing, so that at light 
quark masses the total quenched chiral loop contribution to the 
$\Delta$ mass is repulsive, whereas it is attractive in full QCD. 
%

It is now straightforward to fit the quenched lattice data with the form:
\begin{equation}
\tilde{M}_B = \tilde{\alpha}_B + \tilde{\beta}_B \mpi^2 + 
\tilde{\Sigma}_B (\mpi, \Lambda) .
\label{Eq:quenchfit}
\end{equation}
Once again the linear part describes how the mass of the pion-cloud 
source varies with quark mass. This form includes the expected 
behaviour of HQET where the $\pi$ and $\eta'$ loop contributions 
are suppressed.
Since, as discussed earlier, the meson--baryon vertices are 
characterised by the source distribution, which is argued to be 
similar in quenched and full QCD,  we take all vertices 
to have the same momentum dependence --- i.e. a common dipole 
mass, $\Lambda$. With this parameter fixed there are just two 
free parameters, $\tilde{\alpha}$ and $\tilde{\beta}$, to fit the 
quenched data for each baryon. 

As described in Ref.~\cite{Leinweber:2001ac} we replace the 
continuum integral over the intermediate pion momentum by a 
discrete sum over the pion momenta available on the lattice, 
thus encapsulating finite lattice spacing and volume 
artifacts.

\begin{figure}[tbh]
\begin{center}
{\epsfig{file=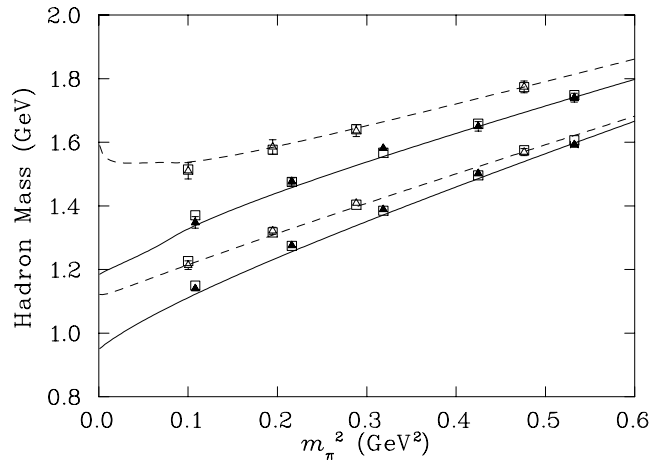, width=\figsize, angle=90}}
\caption{Fit (open squares) to lattice data (\protect\cite{Bernard:2001av}: 
Quenched $\vartriangle$, Dynamical $\blacktriangle$) with adjusted 
self-energy expressions accounting for finite volume 
and lattice spacing artifacts.
The continuum limit of quenched (dashed lines) and dynamical (solid lines) 
are shown. The lower curves are for $N$ and upper for $\Delta$.}
\label{fig:fqFit}
\end{center}
\end{figure}
\begin{ruledtabular}
\begin{table}[bth]
\begin{center}
\begin{tabular}{lcccc}
          & $\alpha_N$    & $\beta_N$       & $\alpha_\Delta$     & $\beta_\Delta$     \\
\hline
Full      & $1.24(2)$     & $0.92(5)$       & $1.43(3)$           & $0.75(8)$          \\
Quenched  & $1.23(2)$     & $0.85(6)$       & $1.45(4)$           & $0.71(11)$         \\
\end{tabular}
\caption{Best fit parameters for both full and quenched 
data sets for dipole mass, $\Lambda=0.8\,{\rm GeV}$. All masses in are GeV.
\label{tab:fitparams}}
\end{center}
\end{table}
\end{ruledtabular}
The lattice data used for this analysis comes from a recent paper 
by Bernard {\em et al.} \cite{Bernard:2001av}. These results are 
obtained using an improved Kogut-Susskind quark action, which is 
known to give good scaling properties \cite{Bernard:1999xx}. 
Unlike the standard Wilson fermion action, masses determined 
at finite lattice spacing are excellent estimates of the continuum 
limit results.
The physical scale of both full and quenched data sets has been 
set via a variant of the Sommer scale \cite{Bernard:2001av}.
This procedure, based on the static-quark potential where chiral 
corrections are negligible \cite{tocome}, provides a 
self-consistent determination of the scale for both simulations.

Fits of the the form for full QCD, Eq.~\ref{Eq:fullfit}, and 
quenched QCD, Eq.~\ref{Eq:quenchfit}, are shown in 
Fig.~\ref{fig:fqFit}. In fitting to data we choose a 
dipole mass of $\Lambda=0.8\,{\rm GeV}$, which has been 
optimised to highlight the remarkably similar behaviour of 
the pion-cloud source in both quenched and dynamical simulations. 
Phenomenologically, this agrees with quite general expectations 
that it should be somewhat smaller than that for the axial form 
factor \cite{Thomas:2001kw,Guichon:1983zk,Thomas:1989tv}. 
The parameters obtained from our fits are shown in 
Table~\ref{tab:fitparams}.

We highlight the fact that the best fit parameters, $\alpha$ and 
$\beta$, obtained from both the quenched and full simulations, 
agree within errors. This suggests that the quark mass behaviour 
of the pion-cloud source is quite similar in full and quenched 
QCD, lending support to the hypothesis made in this analysis. 
This leads one to conclude that the dominant effects of 
quenching can be attributed to the first order meson loop corrections.


We have investigated the quark mass dependence of the $N$ and $\Delta$ 
masses within the quenched approximation. The leading chiral behaviour 
of hadron masses in quenched QCD is known to differ from the full theory. 
This knowledge has been used to guide the construction of a 
functional form which both reproduces this correct chiral structure, 
is consistent with current lattice simulations and encompasses the 
HQET properties. The success of this 
method in the quenched case further verifies the importance of 
including meson-induced self-energies when extrapolating lattice results.

We find that, although the quenched approximation gives rise 
to more singular behaviour in the chiral limit, 
these contributions are quickly suppressed with increasing quark mass.
In the nucleon, the effects of quenching reduce the amount of curvature 
expected as lighter quark masses are simulated. 
In contrast, for the $\Delta$ we find some upward curvature of 
the mass in QQCD as the quark mass approaches zero. In addition, the 
$\Delta$--$N$ mass splitting increases to around 400 MeV at the 
physical point. As a consequence of this behaviour, the $\Delta$ 
mass in the quenched approximation is expected to differ from the 
physical mass by approximately $25\%$.

Our calculations suggest that the one-loop meson graphs which generate
the leading and next-to-leading non-analytic behaviour are the primary
difference between baryon masses in quenched and full QCD. 
Thus, rather than quenched lattice QCD being regarded as an 
uncontrolled approximation, we are able to make a quantitative estimate of 
errors over a range of quark mass. 
It is vital to investigate the assumptions made with further dynamical 
fermion simulations at low quark masses to test the extent to which 
the presented results hold. 
Nevertheless, this discovery represents a remarkable 
step forward in relating lattice QCD to observed hadronic properties.


We would like to thank S.~Sharpe for numerous enlightening discussions 
concerning Q$\chi$PT as well as W.~Detmold, M.~Oettel, A.~Williams and 
J.~Zanotti for helpful conversations. 
This work was supported by the Australian Research Council and the
University of Adelaide.


\begin{thebibliography}{30}


\bibitem{Bowler:1999ae}
{\bf UKQCD} Collaboration, K.~C. Bowler {\em et al.} {\em Phys. Rev.} {\bf D62}
  (2000) 054506.

\bibitem{Kanaya:1998sd}
{\bf CP-PACS} Collaboration, K.~Kanaya {\em et al.} {\em Nucl. Phys. Proc.
  Suppl.} {\bf 73} (1999) 189--191.

\bibitem{Bernard:2001av}
C.~Bernard {\em et al.} {\em Phys. Rev.} {\bf D64} (2001) 054506.

\bibitem{Zanotti:2001yb}
J.~M.~Zanotti {\it et al.} 
``Hadron masses from novel fat-link fermion actions,''
accepted for publication in {\em Phys. Rev.}~{\bf D}, 
{\tt arXiv:hep-lat/0110216}; and in preparation.

\bibitem{Aoki:1999ff}
{\bf CP-PACS} Collaboration, S.~Aoki {\em et al.} {\em Phys.
  Rev.} {\bf D60} (1999) 114508.

\bibitem{Dyson:1952tj}
F.~J. Dyson {\em Phys. Rev.} {\bf 85} (1952)
631--632.

\bibitem{LeGuillou:1990nq}
J.~C. Le~Guillou and J.~Zinn-Justin, eds., {\em Large order behavior of
  perturbation theory}.
\newblock Amsterdam, Netherlands: North-Holland (1990) 580 p.

\bibitem{magmom}
D.~B. Leinweber, D.~H. Lu, and A.~W. Thomas {\em Phys. Rev.} {\bf D60} (1999)
  034014; 
E.~J. Hackett-Jones, D.~B. Leinweber, and A.~W. Thomas {\em Phys. Lett.} {\bf
  B489} (2000) 143.

\bibitem{Hackett-Jones:2000js}
E.~J. Hackett-Jones, D.~B. Leinweber, and A.~W. Thomas {\em Phys. Lett.} {\bf
  B494} (2000) 89--99.

\bibitem{Detmold:2001jb}
W.~Detmold {\em et al.} {\em Phys. Rev. Lett.} {\bf 87} (2001) 172001.

\bibitem{Dunne:2001ip}
G.~V. Dunne, A.~W. Thomas, and S.~V. Wright, 
``Chiral extrapolation: An analogy with effective field theory,'' 
accepted for publication in {\em Phys. Lett.}~{\bf B}, 
{\tt arXiv:hep-th/0110155}.

\bibitem{Leinweber:1999ig}
D.~B. Leinweber, A.~W. Thomas, K.~Tsushima, and S.~V. Wright {\em Phys. Rev.}
  {\bf D61} (2000) 074502.

\bibitem{Bernard:1992mk}
C.~W. Bernard and M.~F.~L. Golterman {\em Phys. Rev.} {\bf D46} (1992)
  853--857.

\bibitem{Labrenz:1996jy}
J.~N. Labrenz and S.~R. Sharpe {\em Phys. Rev.} {\bf D54} (1996) 4595--4608.

\bibitem{Sharpe:2000bc}
S.~R.~Sharpe and N.~Shoresh, Phys.\ Rev.\ D {\bf 62} (2000) 094503.

\bibitem{Sharpe:2001fh}
S.~R.~Sharpe and N.~Shoresh, Phys.\ Rev.\ D {\bf 64} (2001) 114510.

\bibitem{Leinweber:1993hj}
D.~B. Leinweber and T.~D. Cohen {\em Phys. Rev.} {\bf D47} (1993) 2147--2150.

\bibitem{Leinweber:1991dv}
D.~B. Leinweber, R.~M. Woloshyn, and T.~Draper {\em Phys. Rev.} {\bf D43}
  (1991)
1659--1678.

\bibitem{tocome}
R.~D. Young, D.~B. Leinweber, A.~W. Thomas, and S.~V. Wright, in preparation.

\bibitem{Isgur:1999jv}
N.~Isgur {\em Phys. Rev.} {\bf D62} (2000) 054026.

\bibitem{Bernard:1995dp}
V.~Bernard, N.~Kaiser, and U.-G. Meissner {\em Int. J. Mod. Phys.} {\bf E4}
  (1995) 193--346.

\bibitem{Sharpe:1992ft}
S.~R. Sharpe {\em Phys. Rev.} {\bf D46} (1992) 3146--3168.

\bibitem{Leinweber:2001ac}
D.~B. Leinweber, A.~W. Thomas, K.~Tsushima, and S.~V. Wright {\em Phys. Rev.}
  {\bf D64} (2001) 094502.

\bibitem{Bernard:1999xx}
{\bf MILC} Collaboration, C.~W. Bernard {\em et al.} {\em Phys. Rev.} {\bf D61}
  (2000) 111502.

\bibitem{Thomas:2001kw}
A.~W. Thomas and W.~Weise, {\em The Structure of the Nucleon}.
\newblock Wiley-VCH, Berlin, 2001.

\bibitem{Guichon:1983zk}
P.~A.~M. Guichon, G.~A. Miller, and A.~W. Thomas {\em Phys. Lett.} {\bf B124}
  (1983) 109.

\bibitem{Thomas:1989tv}
A.~W. Thomas and K.~Holinde {\em Phys. Rev. Lett.} {\bf 63} (1989)
2025--2027.

\end{thebibliography}
\end{document}